\documentclass[a4paper]{jpconf}
\usepackage{graphicx}
\begin{document}
\title{VIP2 in LNGS - Testing the Pauli Exclusion Principle for electrons with high sensitivity}

\author{J. Marton$^{1}$, A. Pichler$^{1}$,  A. Amirkhani$^{4}$, S. Bartalucci$^{2}$, M. Bazzi$^{2}$, S. Bertolucci$^{6}$, M. Bragadireanu$^{2,3}$, M. Cargnelli$^{1}$, A. Clozza$^{2}$, C. Curceanu$^{2,3,9}$, R. Del Grande$^{2}$, L. De Paolis$^{2}$,  J.-P. Egger$^{7}$, C. Fiorini$^{4}$, C. Guaraldo$^{2}$, M. Iliescu$^{2}$, M. Laubenstein$^{8}$, E. Milotti$^{5}$, M. Milucci$^{2}$, D. Pietreanu$^{2,3}$, K. Piscicchia$^{9,2}$, A. Scordo$^{2}$, H. Shi$^{10}$, D. Sirghi$^{2,3}$, F. Sirghi$^{2,3}$, L. Sperandio$^{2}$, O. Vazquez-Doce$^{2,11}$ and J. Zmeskal$^{1}$}
\address{$^{1}$ Stefan Meyer Institute for subatomic physics, Boltzmanngasse 3, 1090 Vienna, Austria}
\address{$^{2}$ INFN, Laboratori Nazionali di Frascati, CP 13, Via E. Fermi 40, I-00044, Frascati (Roma),
Italy}
\address{$^{3}$ “Horia Hulubei National Institute of Physics and Nuclear Engineering, Str. Atomistilor no.
407, P.O. Box MG-6, Bucharest - Magurele, Romania}
\address{$^{4}$ Politecnico di Milano, Dipartimento di Elettronica, Informazione e Bioingegneria and INFN Sezione di Milano, Milano, Italy}
\address{$^{5}$ Dipartimento di Fisica, Universit{\`a}  di Trieste and INFN– Sezione di Trieste, Via Valerio, 2,
I-34127 Trieste, Italy}
\address{$^{6}$ University and INFN Bologna, Via Irnerio 46, I-40126, Bologna, Italy}
\address{$^{7}$ Institut de Physique, Universit{\'e} de Neuch\^{a}tel 1 rue A.-L. Breguet, CH-2000 Neuch\^{a}tel, Switzerland}
\address{$^{8}$ INFN, Laboratori Nazionali del Gran Sasso, S.S. 17/bis, I-67010 Assergi (AQ), Italy}
\address{$^{9}$ Centro Fermi - Museo Storico della Fisica e Centro Studi e Ricerche "Enrico Fermi", Roma, Italy}
\address{$^{10}$ Institut f{\"ü}r Hochenergiephysik, Nikolsdorfergasse 18, 1050 Vienna, Austria}
\address{$^{11}$ Excellence Cluster Universe, Technische Universit{\"ä}t  M{\"ü}nchen, Boltzmannstra{\ss}e 2, D-85748 Garching, Germany}
\ead{johann.marton@oeaw.ac.at} 

\begin{abstract}
The VIP2 (VIolation of the Pauli Exclusion Principle) experiment at the Gran Sasso underground laboratory (LNGS) is searching for possible violations of standard quantum mechanics predictions in atoms at very high sensitivity. We investigate atomic transitions with precision X-ray spectroscopy in order to test the Pauli Exclusion Principle (PEP) and therefore the related spin-statistics theorem. We will present our experimental method for the search for "anomalous" (i.e. Pauli-forbidden) X-ray transitions in copper atoms, produced by "new" electrons, which could have tiny probability to undergo Pauli-forbidden transition to the ground state already occupied by two electrons. We will describe the VIP2 experimental setup, which is taking data at LNGS presently. The goal of VIP2 is to test the PEP for electrons with unprecedented accuracy, down to a limit in the probability that PEP is violated at the level of 10$^{-31}$. We will present current experimental results and discuss implications of a possible violation.
\end{abstract}

\section{Introduction}
The Pauli Exclusion Principle (PEP) was proposed by Wolfgang  Pauli \cite{pauli25} who received the Nobel prize for this finding. It is regarded as a consequence of the spin-statistics theorem. In his Nobel prize lecture Pauli noticed \cite{pauli-nobel} 
{\it we want to stress here a law of Nature which is generally valid, namely, the connection between spin and symmetry class. A half-integer value of the spin quantum number is always connected with antisymmetrical states (exclusion principle), an integer spin with symmetrical states}. 
First the PEP was found for electrons and later extended to all fermions (half-integer spin systems).
Thus the nature is divided into bosonic (symmetric) and fermionic (antisymmetric) systems. {\it Is anything between, with mixed symmetry?} is an obvious question.
Up-to-now no contradiction with the validity of PEP was found. Therefore,  the PEP is extremely successful in explaining many fundamental features of nature. Here some examples:  periodic table of the elements, the existence of neutron stars and the stability of matter \cite{lieb76,Cern-Courier18}. Therefore the spin plays a decisive role, dividing nature in half-integer and integer particles and systems, i.e. in fermions and bosons respectively.
Nevertheless, to probe the PEP validity with utmost sensitivity is extremely important because of its role in foundations of quantum physics. Even a tiny violation would point to new physics. The combination of high sensitivity X-ray spectroscopy and the large background suppression in an underground laboratory like LNGS is employed to test PEP in the VIP2 experiment.

\section{Experimental tests  of PEP}

In the past searches for both PEP-violating transitions and PEP-forbidden system were carried
out. Such investigations were performed with atoms and nuclei but also astrophysical bounds on
the PEP were searched. The accuracy of tests of PEP for electrons has gradually increased over
time (see table 1).
The quantity $\beta^{2}$/2 - widely used in the literature - gives the upper limit of PEP
violation and can be traced back to a model introduced by Ignatiev and Kuzmin \cite{Ignatiev87}. We are using
this quantity to enable a comparison of the results employing different approaches. The Messiah
and Greenberg superselection rule \cite{messiah64} constrains the possible transitions as it forbids symmetry
changes in stable systems. Experiments investigating stable systems reached extremely low
upper limits for PEP violation \cite{bernabei09,back04}. Regarding atomic transitions the first experimental test of PEP for electrons can be traced back to a pioneering experiment by Goldhaber and Scharff-Goldhaber more than half a century ago. In that experiment new electrons from beta decay
were used as probes of the validity of PEP. About 40 years later the validity of PEP for electrons
was examined in an experiment by Ramberg an Snow using a current to bring new electrons
into the system resulted in an upper limit for PEP violation in the order of 10$^{-26}$. The VIP2
experiment follows the Ramberg-Snow approach with largely improved instrumentation and in
a low background LNGS environment.

\begin{table}[h]
\caption{\label{tabone}Upper bounds on $\beta^{2}$/2 from experimental tests of the PEP validity searching for PEP-forbidden atomic transitions. Only experiments in accordance with the Messiah-Greenberg superselection rule are given. The VIP experiment repeatedly succeeded in improving the bound by optimization of the experimental setup \cite{pichler18}.} 

\begin{center}
\lineup
\begin{tabular}{*{5}{l}}
\br                              
Year&experiment& Supply of new electrons & $\beta^{2}$/2 &reference\cr 
\mr
1948 & Goldhaber/Scharff-Goldhaber       & beta-rays         &  3x10$^{-3}$ &  \cite{goldhaber48} \cr
1990 & Ramberg/Snow                             & electric current & 1.7x10$^{-26}$ & \cite{ramberg90} \cr
2006 & VIP                           & electric current & 4.5x10$^{-28}$ & \cite{bartalucci06} \cr
2011 & VIP2                         & electric current & 4.7x10$^{-29}$ & \cite{curceanu11} \cr
2018 & VIP2 & electric current &3.4x10$^{-29}$ & \cite{shi18,Shi18}  \cr
\br
\end{tabular}
\end{center}
\end{table}

\section{The VIP2 experiment at LNGS}

\subsection{Experimental concept of VIP2}

The VIP2 experiment at the Laboratori Nazionali del Gran Sasso (LNGS) searches for Pauli-forbidden X-ray transitions in copper at the utmost sensivity (see fig.1). If an atomic transition from the 2p state to the already filled 1s ground state (K$_{\alpha}$ transition)  takes place - thus violating PEP - then the transition energy is lower by about 300 eV (i.e.  the transition energy for the
PEP violating transition is 7747 eV instead of 8048 eV) which can be determined by modern
X-ray spectroscopy employing solid state X-ray detectors. The new electrons as probes of PEP
are introduced by an electric current circulating through a copper foil. Therefore, periods with
current (possible PEP-violating transitions) and without current (no PEP violating transitions)
are alternating in VIP2.

\begin{figure}
\begin{center}
\includegraphics[width=15cm]{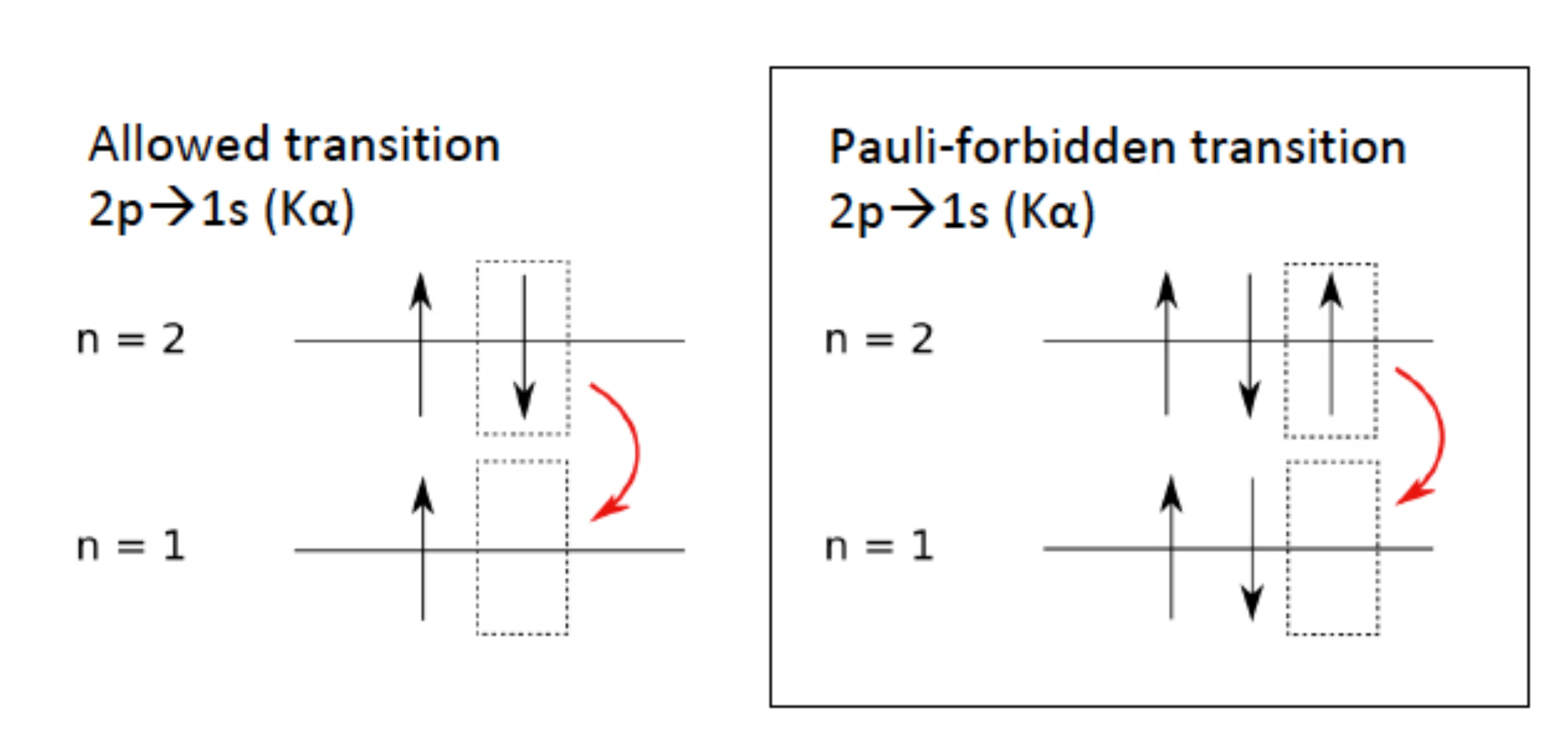}
\end{center}
\caption{\label{figure1}The VIP2 experiment is searching for Pauli-forbidden the atomic transition to the already filled groundstate n=1 (K$_{\alpha}$) displayed in the right picture.}
\end{figure}

\subsection{VIP2 setup}

The VIP experiment used charge-coupled devices (CCDs) for the X-ray detection. CCDs provide
excellent energy resolution but the background reduction is limited because of the lack of timing
information. Subsequently the VIP2 experiment employs Silicon Drift Detectors (SDDs) (see fig.2). These semiconductor detectors are ideally suited for soft X-ray spectroscopy combining large active area, small capacity, arrangements in arrays and provide excellent energy resolution and timing capability. The
X-rays generate free electrons, which drift to the anode with an applied electric field. The energy spectrum of the X-rays can be deduced from the electrons arriving at the anode. This current is amplified and read out by an ADC. 
The timing performance is
sufficient (below 1$\mu$s) for anticoincidence with scintillation detectors in order to suppress actively
background. Their larger depletion depth leads to a higher detection efficiency of X-rays in the
region of interest.  The timing capability of SDDs  enables
the use of an active shielding with scintillators against external radiation. For this purpose, 32
scintillator bars read out by silicon photomultipliers are installed around
the SDDs. In addition to the passive shielding around the VIP2 apparatus their veto signal helps to reduce  further  the background.

\begin{figure}
\begin{center}
\includegraphics[width=15cm]{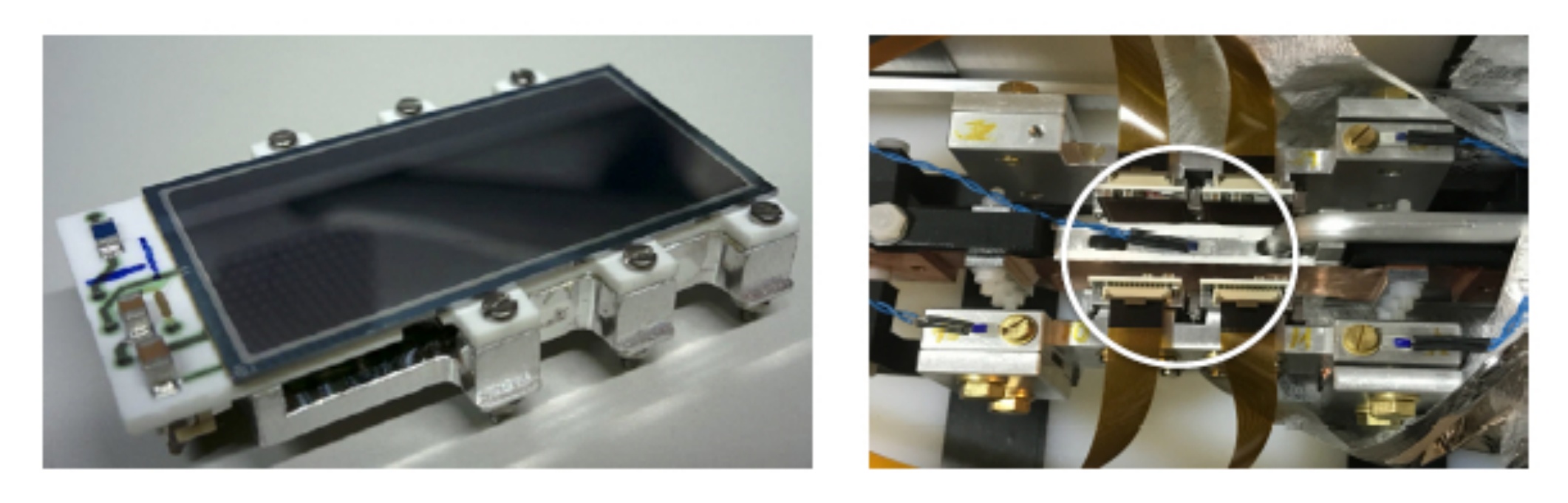}
\end{center}
\caption{\label{figure2}SDD detector (left) consisting of an array of 2x4 SDDs and the arrangement in the vacuum box of VIP2 (right). The white circle indicates the position of the X-ray detectors positioned on both sides of the water-cooled copper target conducting a current of about 100A.}
\end{figure}

\subsection{VIP2 Shielding} 

The shielding of the VIP2 apparatus employs active and passive shielding. The active shielding is using plastic bars readout by silicon photomultipliers. The efficiency of the active shielding depends on the lower threshold of the scintillation detectors. Presently tests and optimization of the active shielding system is in progress. The passive shielding consists of copper and lead blocks surrounding the outer part of the VIP2 apparatus. This method of passive shielding was successfully employed in the VIP-CCD experiment.

\section{Measurements and Results}

Recently the VIP2 experimental setup was upgraded. In 2018 the new Silicon Drift Detectors \cite{quaglia15} (see fig.2) were mounted in the VIP2 setup in the
underground laboratory Gran Sasso (LNGS). 
These detectors consist of 4 arrays with 8 single cell SDDs per array. Each square cell has 8 x 8
mm$^{2}$ surface area. In total the active detector arrea is around 20 cm$^{2}$. Compared to the formerly employed SDDs the active area is increased by a factor larger than 3. Therefore, the detection
efficiency for possible photons from Pauli Exclusion Principle (PEP) violating transitions is
increased by a factor greater than 3. An energy resolution of 150 eV -160 eV (FWHM) at 6 keV
could be verified for all channels. The experiment is taking data with and without current under
stable conditions with the new configuration (see fig.3). Furthermore, Monte Carlo simulations
conducted with GEANT 4, which have the environmental gamma radiation spectrum as input,
verified the background rate detected with this new configuration. The progress of the VIP2 experiment has  been reported in \cite{pichler18, Shi18, marton18, Curceanu2017, curceanu17}. In 2016 we collected data in a time period of $\sim$70 without current and $\sim$40 days with 100 A current. In fig.3 a typical X-ray spectrum measured with the SDD detector system is displayed.

\begin{figure}
\begin{center}
\includegraphics[width=15cm]{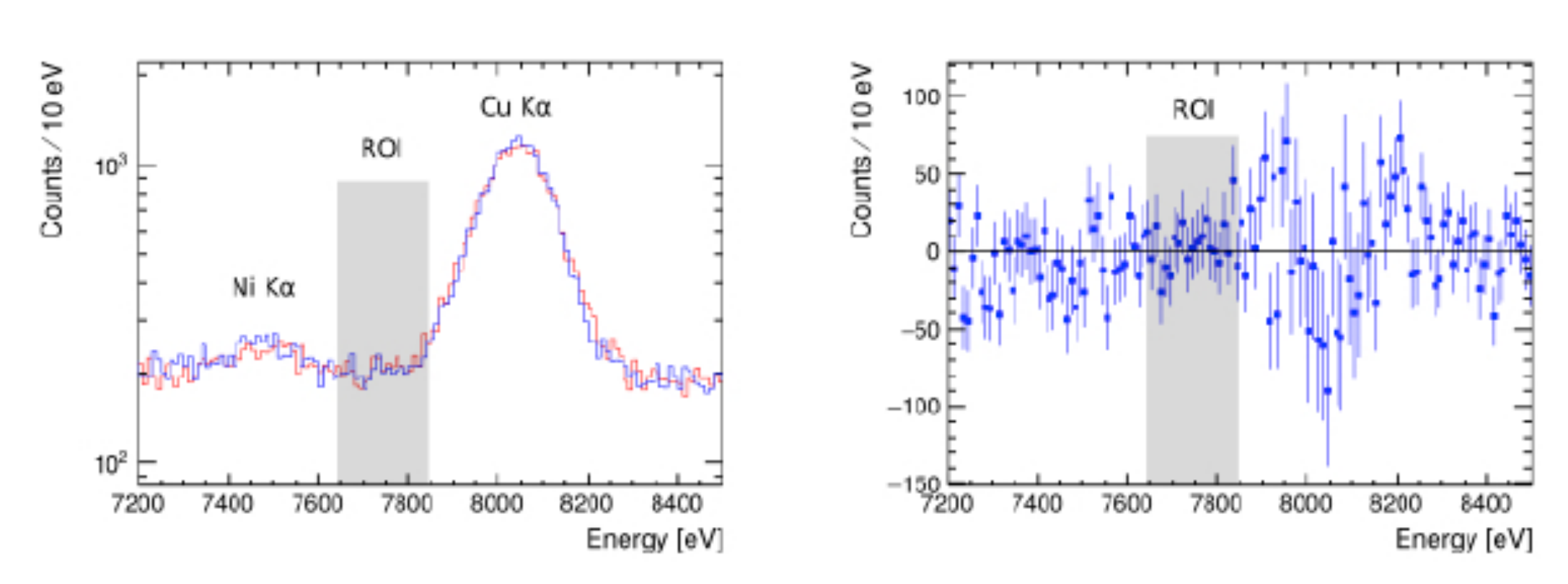}
\end{center}
\caption{\label{figure3}The X-ray spectrum measured with current and without current in VIP2 (left). The region-of interest (ROI) is shaded. After subtraction of the background events the residual X-ray spectrum is shown in the right panel.}
\end{figure}


\section{Summary and Outlook}
The experimental program for testing a possible PEP violation for electrons made great progress so far. The use of a new type of SDDs as X-ray detectors will further enhance the sensitivity by providing larger sensitive area. 
Concerning the reduction of the X-ray background we installed
a part of the passive shielding with copper and lead which will be completed in 2019. Given a
running time of 3 years and alternating measurement with and without current we might either
find a tiny violation or we can lower the upper limit of PEP violation by about two orders of
magnitude (see fig.4). On the theoretical side studies are made to investigate the concept of
new electrons provided by a current in the copper bulk material \cite{milotti18 }.

\begin{figure}
\begin{center}
\includegraphics[width=15cm]{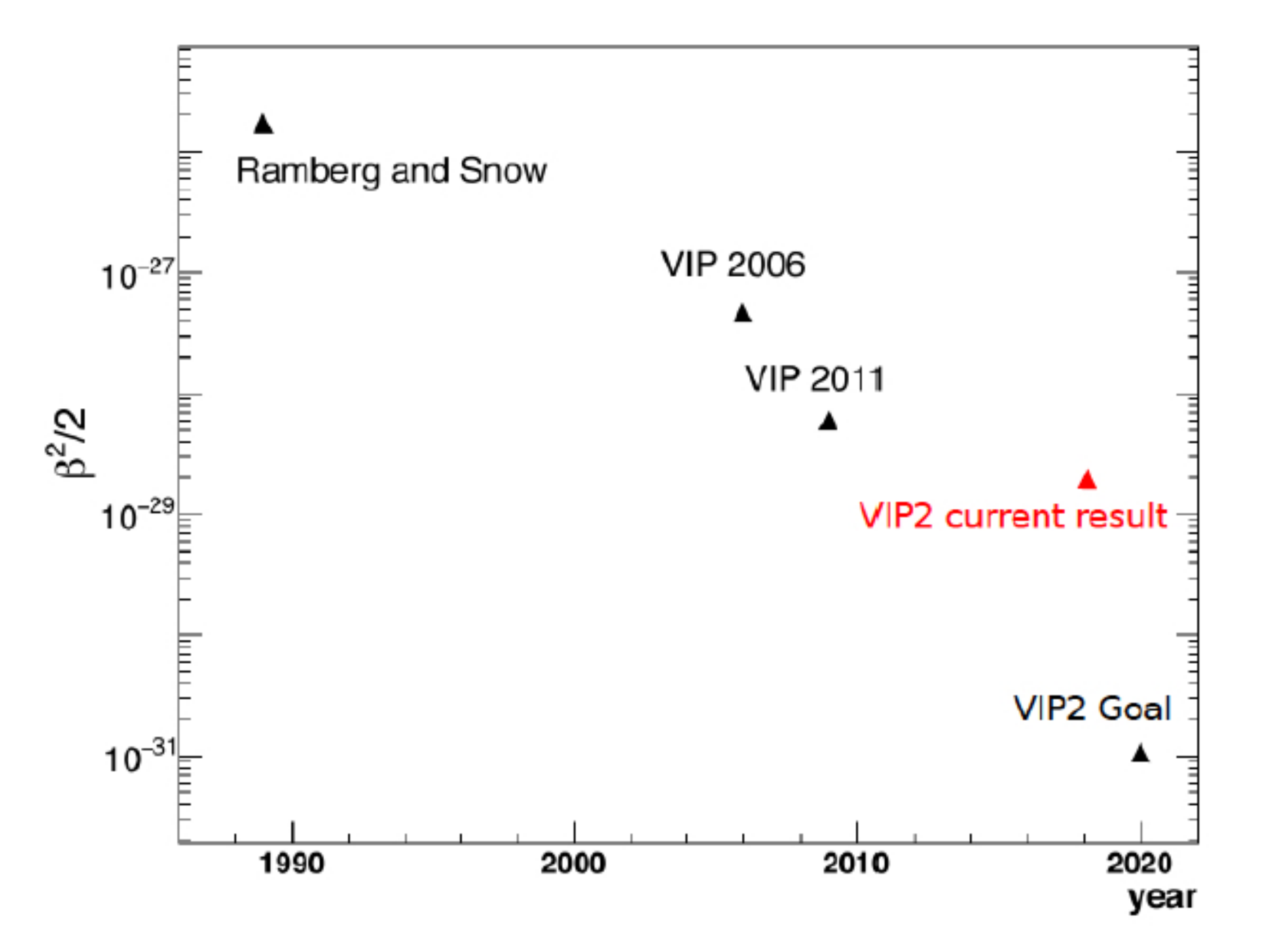}
\end{center}
\caption{\label{figure4}Progress in lowering the upper limit of PEP violation in atomic transitions (respecting
the Messiah-Greenberg superselection rule) in the last 20 years. The goal of the new VIP2
experiment with a running time of about 3 year is indicated.}
\end{figure}

\ack
We thank H. Schneider, L. Stohwasser, and D. Pristauz-Telsnigg from Stefan-Meyer-Institut
for their fundamental contribution in designing and building the VIP2 setup. We acknowledge
the very important assistance of the INFN-LNGS laboratory. The support from the EU COST
Action CA 15220 and of the EU FET project TEQ (grant agreement 766900) is gratefully
acknowledged. We thank the Austrian Science Foundation (FWF) which supports the VIP2
project with the grants P25529-N20, project P 30635-N36 and W1252-N27 (doctoral college
particles and interactions) and Centro Fermi for the grant Problemi Aperti nella Meccanica
Quantistica. Furthermore, these studies were made possible through the support of a grant from
the Foundational Questions Institute, FOXi (and a grant from the John Templeton Foundation
(ID 58158). The opinions expressed in this publication are those of the authors and do not
necessarily respect the views of the John Templeton Foundation.


\section*{References}

\begin{thebibliography}{10}


\bibitem{pauli25} Pauli W 1925 {\it Z. Phys.} {\bf 31} 765

\bibitem{pauli-nobel} Pauli W 1945 Wolfgang Pauli – Nobel Lecture

\bibitem{lieb76} Lieb E. H. 1976 The stability of matter  {\it Rev. Mod. Phys.} {\bf 48} 553

\bibitem{Cern-Courier18} Curceanu C, Marton J,  Budker D, Milotti E and Hall E Putting the
Pauli exclusion principle on trial {\it Cern Courier} 2018

\bibitem{Ignatiev87} Ignatiev A Yu and Kuzmin V A 1987 {\it Yad. Fiz.} {\bf 46} 786

\bibitem{messiah64} Messiah A M I and Greenberg O W 1964 {\it Phys. Rev. B}  {\bf 136} 248

\bibitem{bernabei09} Bernabei R  {\it et al} 2009 New search for processes violating the Pauli exclusion principle in sodium and in iodine
 {\it Eur. Phys. J. C}  {\bf 62} 327

\bibitem{back04} Back O.  {\it et al} 2004 New experimental limits on violations of the Pauli exclusion principle obtained with the
Borexino Counting Test Facility  {\it European Physical Journal} C {\bf 37} 421, 10.1140/epjc/s2004-01991-1.

\bibitem{pichler18} Pichler A 2018 Test of the Pauli Exclusion Principle for electrons in the Gran Sasso underground laboratory
PhD Thesis, TU Vienna to be published

\bibitem{goldhaber48} Goldhaber M and Scharff-Goldhaber G 1948  Identification of Beta-Rays with
Atomic Electrons. {\it Physical Review} {\bf 73(12)} 472

\bibitem{ramberg90}
Ramberg E and Snow G A 1990 Experimental limit on a small violation of the Pauli principle {\it Physics Letters B} {\bf 238(2)} 438

\bibitem{bartalucci06} Bartalucci S  {\it et al} 2006 {\it  Phys. Lett. B} {\bf 641} 18

\bibitem{curceanu11} Curceanu C  {\it et al} 2011 Experimental tests of quantum mechanics: Pauli Exclusion Principle Violation (the
VIP experiment) and future perspectives  {\it Journal of Physics: Conference Series}  {\bf 306(1)} 012036

\bibitem{shi18} Shi H  {\it et al} 2018 Experimental search for the violation of Pauli Exclusion Principle {\it Eur. Phys. J C} {\bf 78} 319. https://doi.org/10.1140/epjc/s10052-018-5802-4

\bibitem{Shi18} Shi H  {\it et al}  2018 Search for the violation of Pauli Exclusion Principle at LNGS {\it EPJ Web of Conferences} {\bf volume 182} 02118


\bibitem{quaglia15} Quaglia, R  {\it et al}  2015 Silicon Drift Detectors and CUBE preamplifiers for high-resolution X-ray spectroscopy  {\it IEEE Transactions on Nuclear Science} {\bf 62 (1)} 221

\bibitem{marton18} Marton J  {\it et al}  2018 Underground Test of Quantum Mechanics: The VIP2 Experiment Quantum Foundations, Probability and Information 155


\bibitem{Curceanu2017}
Curceanu C {\it et al} 2017 {\it Entropy} {\bf 19 (7)} 300 



\bibitem{curceanu17}  Curceanu C  {\it et al}  2017 Quantum mechanics under X-rays in the Gran Sasso underground laboratory  {\it Int. J.
Quant. Inf.} {\bf15} 1740004 doi:10.1142/S0219749917400044





\bibitem{milotti18 } Milotti E  {\it et al} 2018  On
the Importance of Electron Diffusion in a Bulk-Matter Test of the Pauli Exclusion Principle {\it Entropy} {\bf 20(7)}  515


\end{thebibliography}
\end{document}